\documentclass[a4paper,11pt]{article}
\pdfoutput=1 

\usepackage{jinstpub} 
\graphicspath{{.},{./pictures/}}
\usepackage{tabularx}
\usepackage{siunitx}

\title{Geant4 Systematic Study of the FRACAS Apparatus for Hadrontherapy Cross Section Measurements}

\author[1]{E.~Barlerin,\note{Corresponding author.}}
\author{S. Salvador{\href{https://orcid.org/0000-0003-3444-7807}{\includegraphics[height = 2.ex]{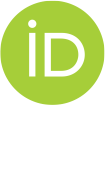}}},}
\author{M.~Labalme}
\affiliation{Laboratoire de Physique Corpusculaire de Caen, Normandie Univ, ENSICAEN, UNICAEN, CNRS/IN2P3, 14000 Caen, France}
\emailAdd{barlerin@lpccaen.in2p3.fr}

\abstract{
In this work, we report on systematic Monte Carlo (MC) studies for the FRACAS apparatus, a large acceptance mass spectrometer that will be used to measure the fragmentation cross sections of $^{12}$C ions for hadrontherapy. The apparatus placed, in a 100~mbar reaction chamber, will be made of a beam monitor, trackers surrounding a magnet and a Time-Of-Flight (TOF) wall. In order to determine the required performances of the trackers, Geant4 simulations of the whole system and in-house developed algorithms were used. While keeping the beam monitor and TOF-wall positions fixed, the effects of the tracker positions and spatial resolutions on the trajectory reconstruction and mass identification efficiencies have been extracted. An optimal configuration was found where the upstream trackers should be located 6~cm away from the target and spaced 4~cm apart whereas their spatial resolutions should be close to 100~\textmu m. The positions of the downstream trackers will have to be changed according to the beam energy to preserve high identification efficiencies. Their spatial resolutions, even though of a lesser importance compared to the upstream trackers, should be around 1~mm or better. In this optimal configuration, an overall fragment identification efficiency above 90\% has been obtained for beam energies ranging from 100 to 400~MeV/nucleon.}

 
\arxivnumber{2009.09839} 

\begin{document}
\bibliographystyle{unsrt}
\maketitle
\flushbottom

\section{Introduction}

During a hadrontherapy treatment, nuclear interactions between the beam ions and the human tissues can occur. These interactions lead to a reduction of the primary beam intensity and to the creation of lighter and longer range fragments resulting in a mixed radiation field. An accurate knowledge of those nuclear interaction processes through their double differential cross sections is then crucial to precisely control the dose deposited in the tumor and the surrounding healthy tissues~\cite{Bohlen2010, Braunn2013, DeNapoli2012}. Although different experiments have already been performed to obtain those cross sections for a $^{12}$C beam below 100~MeV/nucleon on different targets~\cite{dudouet2013, dudouet2014, PLESKAC2012130}, double differential cross sections are still scarce for beam energies between 100 and 400~MeV/nucleon.
   
The FRAgmentation of CArbon and cross Sections (FRACAS) large acceptance mass spectrometer under construction, will be used in the future Archade project\footnote{Advanced Resource Center for HADrontherapy in Europe, Caen, France} to measure the double differential fragmentation cross sections on target of medical interest such as C, H, O, N and Ca. Among the different measurements needed to extract the cross sections such as the fragment kinetic energies and angle of emissions, the identification of the fragment remains crucial. In a mass spectrometer, the particle identification is usually two fold: a first step to extract the charge of the fragment and a second step to identify its mass. In our case, the charge identification will be done by means of a beam monitor and a TOF wall using a $\Delta$E---TOF method~\cite{Toppi2016}. The particle mass identification will be obtained through its magnetic rigidity measured with a set of tracking detectors (also referred as trackers) associated to a large acceptance deflecting magnet. The use of these two simple methods in the case of a mass spectrometer has been proven to be the most accurate identification method for ions with kinetic energies in the range from 150~MeV/nucleon to 400~MeV/nucleon~\cite{PLESKAC2012130,salvador2013simulation} considering the high probability of particle fragmentation in thick calorimeter-like detectors.

In order to optimise the fragment identification efficiencies which rely on the accuracy of the reconstruction of the fragment trajectories, the influence of the tracker properties were studied. Several Geant4 MC simulations for different beam energies were carried out by varying the tracker positions and spatial resolutions. The data were analysed using reconstruction and identification algorithms developed in-house to obtain the trajectory and identification efficiencies of the fragments that led to an optimal configuration of the detection system.

In this paper, the setup of FRACAS is detailed and the MC simulations performed are described along with the reconstruction and identification algorithms. Finally the results of the influence of the tracker positions and spatial resolutions on the fragment identifications are presented and discussed as well as the choice for the most appropriate tracking technologies.

Considering that the tracker technologies were not yet chosen, the combinatorial background on tracks generated by their intrinsic behaviour such as the background noise and read-out type (e.g. pixels or stripped anodes) or even their detection efficiencies were not taken into account. The results are presented as ideal values achieved using the best possible experimental conditions.

\section{Materials and methods}

\subsection{FRACAS setup}
\begin{figure}[!ht]
    \begin{center}
        \includegraphics[width=0.99\linewidth]{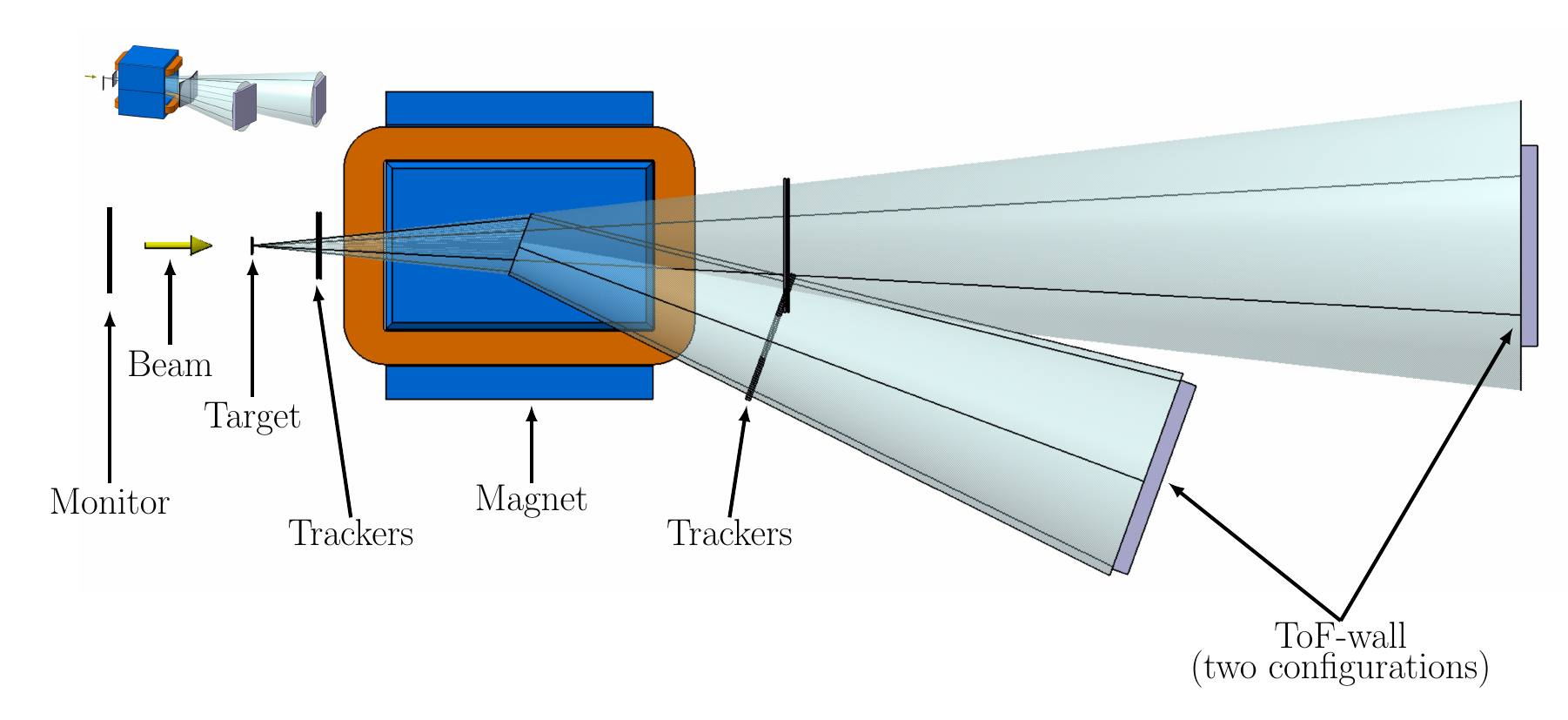}
    \end{center}
    \caption{\label{fig:FRACAS}Sketch of the FRACAS mass spectrometer showing all the detection elements. The TOF-wall is shown in two different configurations depending on the incident beam energy.}
\end{figure}

Figure~\ref{fig:FRACAS} shows a schematic view of the FRACAS experimental setup. It will be composed of a Beam Monitor (BM) located in front of the target, trackers surrounding a magnet and a scintillating detector wall for TOF and energy loss measurements ($\Delta$E). Based on previous simulation studies, the detection system, except for the magnet, will be placed in a large reaction chamber where a pressure of 100~mbar will be made. This pressure was found as a compromise between the reduction of particle scattering in air for long travelling paths and the complexity of the pumping system.

The BM will consist of a multi-stage Parallel Plate Avalanche Counter (PPAC, \cite{BRESKIN1977609}) operated at low isobutane (iC$_{4}$H$_{10}$) pressure. One stage with a thin gap will be used for timing purposes to give the \textit{start} time of the TOF measurement while two other stages with thicker gaps and stripped anodes will be used to extract the beam position and shape with a spatial resolution expected to be below 100~\textmu m in both directions.

The TOF-wall, which will provide the \textit{stop} time for the TOF measurements and the energy released by the fragments, will be a modular system involving 384 scintillating detectors that can be arranged in different configurations. Each module will be composed of a 25.4~$\times$~25.4~mm$^2$ and 1.5~mm thick YAP:Ce crystal coupled to a Hamamatsu R11265-200 ultra-bialkali PhotoMultiplier Tube (PMT). A detailed description of the TOF system along with first timing performances can be found in~\cite{Salvador_2020}.

The magnet will be a large acceptance dipole with a magnetic field of 0.7~T in the center of its gap (detailed in section~\ref{section:MC}) providing sufficient mass separation of the fragments of the same charge.

The trackers will measure the fragment interaction positions in order to obtain their trajectories before and after the magnet, referred to as up and downstream trackers, respectively. The upstream trackers should have small active areas due to their proximity with the target and be made of solid state detectors. On the contrary, the downstream trackers would need to have a large active area of at least 50~$\times$~50~cm$^{2}$ in order to detect as many fragments as possible considering the acceptance of the magnet. The most common detectors to achieve large active areas and low material budget are usually gaseous detectors such as MultiWire Proportional Chambers (MWPCs), or Time Projection Chambers (TPCs). The downstream trackers were then considered as gaseous detectors in our study.

\subsection{Monte-Carlo simulations}
\label{section:MC}
The simulations were made using Geant4.10.5.1 with the \emph{FTFP\_BERT\_HP} physics list for nucleus-nucleus interactions and \emph{electromagnetic\_option3} for electromagnetic processes. In the simulations, only the active parts of the detectors were modelled and the air pressure was reduced to 100~mbar.

The two tracking stages of the BM were both modelled as 10~$\times$~10~cm$^{2}$ active surfaces and 7~mm gap volumes of iC$_{4}$H$_{10}$ at 25~mbar. The timing stage of the beam monitor was modelled with the same active surface and gas pressure but with a gap of 1.6~mm.

The target was modelled as a 5~mm long PMMA (C$_{5}$H$_{8}$O$_{2}$) cylinder with a diameter of 5~mm. The positions of the BM and the target were fixed for all the simulations.

The two upstream trackers were both described as 25~$\times$~25~mm$^{2}$ and 200~\textmu m thick volumes of silicon. The two downstream trackers were modelled as 50~$\times$~50~cm$^{2}$ and 8~mm thick volumes of argon at 1~bar. 

The 384 detection modules of the TOF-wall composed of YAP:Ce crystals coupled to PMTs were arranged in 16 rows of 24 modules. 

In this study, we included in post-analysis the coincidence resolving time of the TOF system (BM and TOF-wall) as a normal distribution with a Full-Width-at-Half-Maximum (FWHM) of 300~ps. The energy resolution of the scintillating detectors $R_L$ following eq.~(\ref{EQENERGRES}) was obtained from experimental measurements using $\gamma$ sources.

\begin{equation}\label{EQENERGRES}
    R_{L} = \left( \frac{187}{\sqrt{E_{0}}} + 1.22 \right)\times L,
\end{equation}
with $L$ the scintillation light in equivalent number of photoelectrons and $E_{0}$ the deposited energy in keV.

Quenching behaviours of the scintillating material were included in the function $L$ using eq.~(\ref{eq:quench}) extracted from~\cite{PARLOG2002674} that converts the deposited energy into equivalent scintillation photoelectrons.

\begin{equation}
    \label{eq:quench}
    L = a_{1} \left[ E_{0} \left[ 1 - a_{2} \frac{ A Z^{2} }{ E_{0} } \text{ln} \left( 1 + \frac{ E_{0} }{ a_{2} A Z^{2} } \right) \right] + a_{2} a_{4} A Z^{2} \text{ln} \left( \frac{ E_{0} + a_{2} A Z^{2} }{ a_{3} A + a_{2} A Z^{2} } \right) \right]
\end{equation}
with $a_{1}$ the conversion factor from energy to collected number of photoelectrons, $a_{2}$, $a_{3}$ and $a_{4}$ the quenching factors whose values are given in table \ref{table:PARAMQUENCH}, A, the mass and Z, the atomic number of the interacting fragment.

\begin{table}[!ht]
    \begin{center}
        \caption{\label{table:PARAMQUENCH}Values of the $a_i$ parameters used in eq.~(\ref{eq:quench}).}
        \renewcommand{\arraystretch}{1.1}{
            \begin{tabularx}{0.7\linewidth}{XXXX}
                \hline
                \hline
               $a_{1}$~(a.u.) & $a_{2}$~(a.u.) & $a_{3}$~(Mev/nucleon) & $a_{4}$ \\
                \hline
                { 19.5 }& { 0.71 } & { 3.8 } & { 0.26} \\
                \hline
                \hline
            \end{tabularx}}
    \end{center}
\end{table}

Concerning the simulation of the deflecting magnet, only its iron frame was modeled with a gap of 70~$\times$~38~$\times$~110~cm$^{3}$ . A measured magnetic field map of the ALADIN magnet installed at GSI~\cite{Toppi2016} was integrated in the simulation.

In this study, the goal is to find the optimal detector characteristics to optimise the identification of the beam particle fragments produced in the target. Therefore, only the $^{12}$C ions that fragmented in the target are considered and the fragments produced elsewhere in the apparatus, for instance in the magnet iron frame, are not included in the data analysis.

\begin{figure}[!ht]
 \begin{center}
    \includegraphics[width=0.99\linewidth]{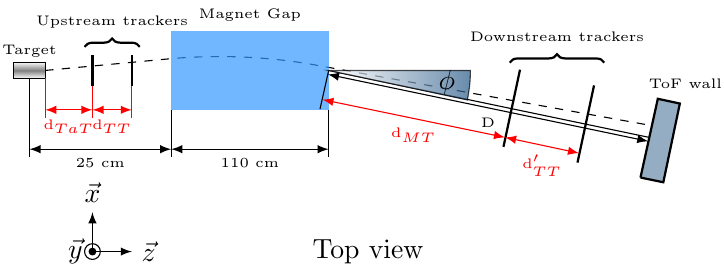}
 \end{center}
 \caption{\label{fig:FRACASABOVE}Simple sketch of the Geant4 simulation of FRACAS showing the different references of the element positions.}
\end{figure}

Figure~\ref{fig:FRACASABOVE} shows a schematic view of the simulation detailing the several position references used in this paper.
The systematic study consisted in varying the positions and spatial resolutions of the trackers. 
The beam monitor, the target and the magnet were kept at the exact same positions for all the simulations. The TOF-wall had different positions for each beam energy, chosen to have the same geometrical efficiency and roughly the same TOF value for the beam ions. Its position was changed at distance $D$ to the magnet and an angle $\phi$ with respect to the beam trajectory so that a $^{12}$C ion from the beam going through the magnetic field impinged in the central TOF-wall module.

The positions of the upstream trackers were defined by the distance of the target to the first upstream tracker ($TaT$) and the distance between the two upstream trackers ($TT_{up}$). Concerning the downstream trackers, their positions were defined by the distance between the exit of the magnet gap and the first downstream tracker ($MT$) and the distance between the two downstream trackers ($TT_{down}$). The trackers were not located closer than 10~cm to the magnet due to the leakage fields.

The simulations were made for four different beam energies: 100, 200, 300 and 400~MeV/nucleon. For each beam energy, multiple simulations were made by varying the up and downstream tracker positions fixing the upstream tracker spatial resolutions to 100~\textmu m and the downstream tracker spatial resolutions to 1~mm in both directions. Table~\ref{table:EFFTRAJRECAP} summarizes the values and ranges of the elements positions used in the simulations for each beam energy.

\begin{table}[!ht]
    \begin{center}
        \caption{\label{table:EFFTRAJRECAP}Values and ranges of the FRACAS element positions evaluated in the Geant4 simulations for the different beam energies.}
        \renewcommand{\arraystretch}{1.1}{
            \begin{tabularx}{1.0\linewidth}{XXXXXXX}
                \hline
                \hline
                Beam energy (MeV/nucleon) & $\phi$ (\textdegree) & $D$ (cm) & $TaT$ (cm) & $TT_{up}$ (cm) & $MT$ (cm) & $TT_{down}$ (cm) \\
                \hline
                { 100 }& { 12.5 } & { 55 } & { $\geq$~5} & { $\leq$~10 } & { $\geq$~10 } & { $\leq$~35} \\
                { 200 }& { 10.5 } & { 85 } & { $\geq$~5} & { $\leq$~10 } & { $\geq$~10 } & { $\leq$~65} \\
                { 300 }& { 8.5 } & { 185 } & { $\geq$~5} & { $\leq$~10 } & { $\geq$~10 } & { $\leq$~165} \\
                { 400 }& { 7.0 } & { 265 } & {  $\geq$~5 } & { $\leq$~10 } & { $\geq$~10 } & { $\leq$~245 }  \\
                \hline
                \hline
            \end{tabularx}}
    \end{center}
\end{table}

Once an optimal configuration was found for all the tracker positions, different values of the tracker spatial resolutions were tested.

To simulate the tracker spatial resolutions, the positions of the particles measured by the detectors were randomly generated in post-analysis following a normal distribution centered on the simulated interaction position and with an FWHM as the spatial resolution. The spatial resolutions were varied between 100~\textmu m to 1.5~mm for the upstream trackers and 200~\textmu m to 3~mm for the downstream trackers. Concerning the downstream trackers, their spatial resolutions in the $x$ and $y$ directions were varied independently as the magnetic field deviates the fragments only in the $x$ direction. 

A last set of simulations was made with the optimal configuration of the tracker positions and spatial resolutions for each beam energy in order to obtain the overall fragment identification efficiencies of the apparatus.

For each simulation, 10$^{7}$ primary $^{12}$C ions impinging on the target were generated to provide sufficient fragment statistics. Table~\ref{table:EFF} gives the number of $^{12}$C ions that fragmented in the target as well as the number of fragments detected in the TOF-wall. The number of fragments that were detected in the same TOF-wall detection module referred as pile-up events are also given. Those events were rejected during the charge identification phase detailed in section~\ref{ssec:ChargID}.

\begin{table}[!ht]
    \begin{center}
        \caption{\label{table:EFF}Number of fragmentation events as well as the number of fragments detected that reached the TOF-wall for each beam energy and 10$^{7}$ primary $^{12}$C ions. Also given is the number of pile-up events detected in the same TOF-wall detection module.}
        \renewcommand{\arraystretch}{1.1}{
            \begin{tabularx}{1.0\linewidth}{XXXX}
                \hline
                \hline
                Beam energy (MeV/nucleon) & \# of fragmentation events ($\times$10$^5$) & \# of fragments detected ($\times$10$^5$)& \# of pile-up events in the TOF-wall \\
                \hline
                { 100 }& { 3.15 } & { 1.82 } & { 530}\\
                { 200 }& { 2.64 } & { 2.59 } & { 750}\\
                { 300 }& { 2.53 } & { 2.79 } & { 710}\\
                { 400 }& { 2.57 } & { 3.09 } & { 760 }\\
                \hline
                \hline
            \end{tabularx}}
    \end{center}
\end{table}

\subsection{Fragment identification}
The identification of a fragment can be decomposed in two parts: the charge reconstruction and the mass reconstruction. However, in order to reconstruct the mass of the fragments, their trajectories must be reconstructed beforehand. The following sections will detail the different processes used in the identification of the fragments.

\subsubsection{Charge identification}
\label{ssec:ChargID}
The charge identification is done using the $\Delta$E---TOF method by plotting the released energy of a charged particle in a material against its time of flight. The different particle charges are then distributed along lines that can be fitted using a simplified version of the Bethe-Bloch formula without radiative corrections:

\begin{equation}
    \Delta E = a \cdot \frac{Z^{2}}{\beta^{2}} \cdot \left[ ln\left(\frac{\beta^{2}b}{1-\beta^{2}} \right) - \beta^{2}\right]
\end{equation}

with $a$ a parameter describing the conversion of the released energy into scintillation photons and $b$ a constant value describing the properties of the YAP crystal given by eq.~(\ref{eq:bmat}) and the semi-empirical formula in eq.~(\ref{eq:ioni}) from~\cite{leo}.
\begin{equation}
\label{eq:bmat}
    b = \frac{2m_{e}c^{2}}{I_{\textrm{YAP}}} = 3486 \qquad\text{with}\quad I_{\textrm{YAP}} = 285~eV
\end{equation}

\begin{equation}
\label{eq:ioni}
    \text{given by~~} \frac{I_{\textrm{YAP}}}{Z_{\textrm{eff}}} = 9.76 + 58.8 \times Z_{\textrm{eff}}^{-1.2} \qquad\text{and}\quad Z_{\textrm{eff}} = 26
\end{equation}

\begin{figure}[!ht]
 \begin{center}
    \includegraphics[width=0.99\linewidth]{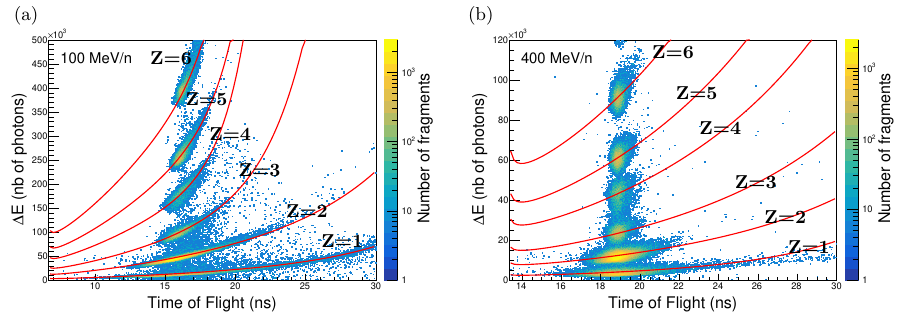}
 \end{center}
 \caption{
 \label{fig:TOF}
 Measured energy released $\Delta$E of the fragments in the TOF-wall detectors converted in scintillation photoelectrons, versus the TOF expressed as the reduced velocity $\beta$ for a $^{12}$C beam at (a) 100~MeV/nucleon and (b) 400~MeV/nucleon. The red curves represent the Bethe-Bloch function with varying Z from 1 to 6.}
\end{figure}

Figure~\ref{fig:TOF} shows examples of the energy released $\Delta$E of the fragments in the TOF-wall detectors, converted in scintillation photoelectrons, versus the time of flight for a $^{12}$C beam at (a) 100~MeV/nucleon and (b) 400~MeV/nucleon. The red curves represent the Bethe-Bloch function with varying $Z$ from 1 to 6 whose parameters $a$ are fitted to the different event populations. The charge of the fragment is then obtained by minimizing the event distance to the closest Bethe-Bloch line using a simple dichotomy algorithm. A more detailed description of this method can be found in~\cite{Toppi2016}.

\subsubsection{Trajectory reconstruction}
Fragments with the same number of charge are separated in a magnetic field according to their magnetic rigidity:

\begin{equation}
    B\rho = 3.10716 \cdot \frac{A}{Z} \cdot \frac{\beta \gamma}{\sin(\theta)}
    \label{Lorentz} 
\end{equation}

 with $\theta$ the angle between the magnetic field and the trajectory of the fragment, $B$ the intensity of the magnetic field, $A$ and $Z$ the number of mass and charge of the fragment, $\beta$ and $\gamma$ the Lorentz factors and 3.10716 the conversion factor between $\frac{m}{q}$ and $\frac{A}{Z}$.
 
 The number of charge $Z$ is already extracted with the $\Delta$E---TOF method and $\beta$ and $\gamma$ are given by the TOF measure. The radius $\rho$ must be obtained through the reconstruction of the trajectories.
 
\begin{figure}[!ht]
    \begin{center}
        \includegraphics[width=0.99\linewidth]{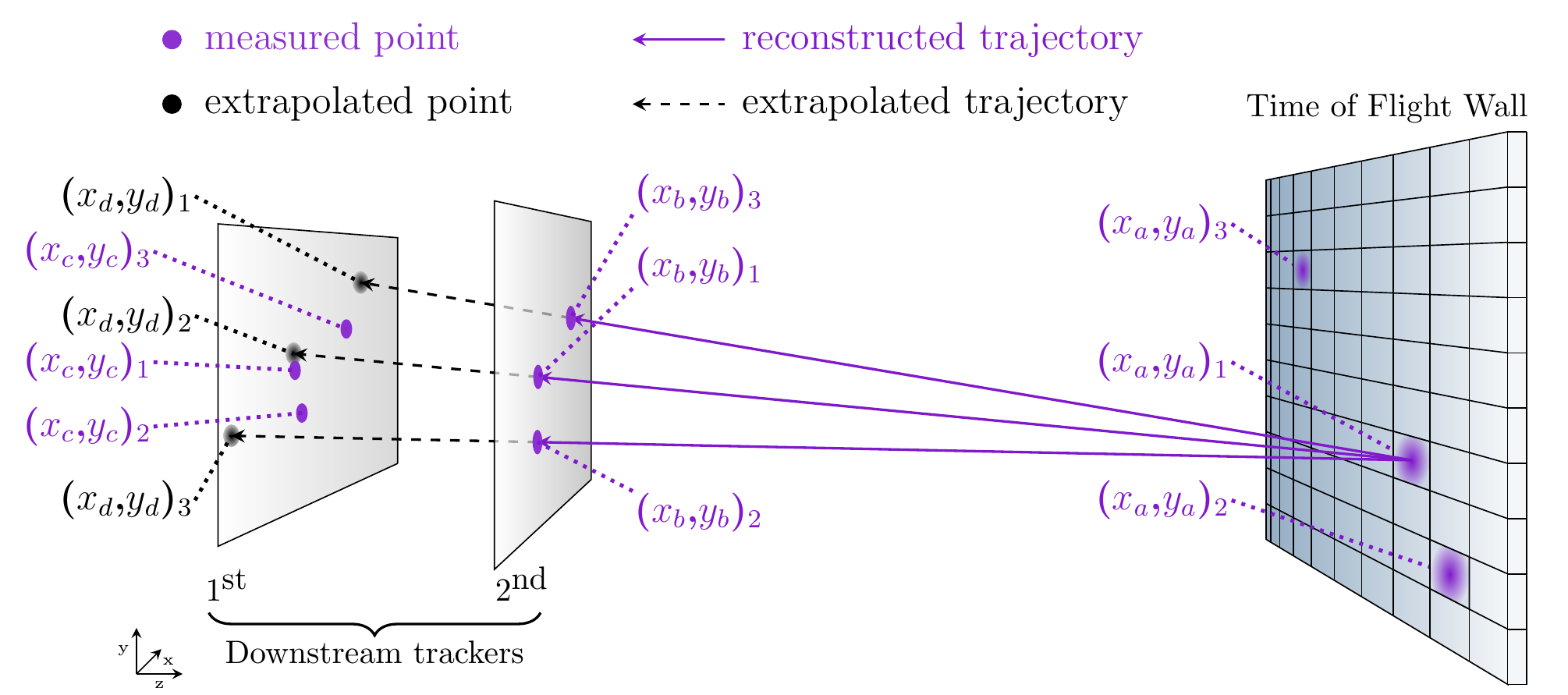}
    \end{center}
    \caption{\label{fig:TRAJRECO}The first steps of the trajectory reconstruction algorithm between the TOF-wall and the downstream trackers. The same can be applied for the reconstruction to the upstream trackers.}
 \end{figure}

 The algorithm used to reconstruct the trajectories of the fragments works as a Kalman filter~\cite{RESCIGNO201434}. It tests all the possible combinations between the positions on each tracker and selects the ones that are most likely to be a trajectory of a fragment. Figure~\ref{fig:TRAJRECO} shows an example of how these steps are computed for the trajectory reconstruction between the TOF-wall and the downstream trackers. It starts from the TOF-wall by taking as a starting point of a trajectory the center of a pixel that has scintillated, ($x_{a}$, $y_{a}$)$_{n}$. It then constructs tracks using this position and the measured positions ($x_{b}$, $y_{b}$)$_{n}$ on the second downstream tracker, and extrapolates those to the first downstream tracker, leading to the ($x_{d}$, $y_{d}$)$_{n}$ points. It was considered that the fragments moved in straight lines in air without scattering. For each measured position ($x_{c}$, $y_{c}$)$_{n}$ on the first downstream tracker, the probability of being part of a trajectory is calculated based on their distance to the extrapolated points ($x_{d}$, $y_{d}$)$_{n}$. 
 
 The steps are repeated to extrapolate to the second and to the first upstream trackers assuming that only the $x$ direction of the fragments are deflected by the magnetic field and that the $y$ coordinates are not affected. The final step is to extrapolate the trajectories to the target, where it is considered that all the trajectories came from its center. The probability for a combination of measured points to be part of a fragment trajectory is the product of the probability computed at each extrapolation step. A threshold is then applied to keep only the most probable combinations. In the case where two reconstructed trajectories shared the same detector position, the one with the highest probability is kept.  The threshold value was set to 10$^{-6}$, giving a trajectory reconstruction efficiency above 90\% in optimal conditions.
 
The radius $\rho$ of the trajectory in the magnetic field is then extracted using the upstream and downstream trajectories and basic trigonometry.
 
\subsubsection{Mass identification}
The mass identification is achieved by plotting the parameters $\rho Z \sin(\theta)$ previously extracted against the reduced velocity $\beta$ of the fragments. The fragments would then distribute along lines given by eq.~(\ref{Lo}) adapted from eq.~(\ref{Lorentz}) according to their number of mass $A$.

\begin{equation}
    f(A) = 3.10716 \frac{A\beta\gamma}{B} 
    \label{Lo}
\end{equation}

Likewise to the charge identification, a simple dichotomy algorithm is used to minimize the distance between a given fragment and each line to associate the fragment with its mass.

\begin{figure}[!ht]
     \begin{center}
        \includegraphics[width=0.8\linewidth]{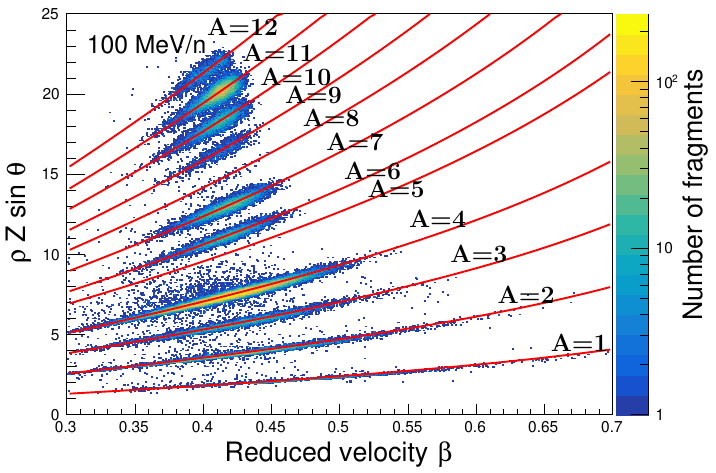}
     \end{center}
     \caption{\label{fig:MASS}Example of a mass identification map showing $\rho Z \sin(\theta)$ as a function of the reduced velocity $\beta$ for a $^{12}$C beam at 100~MeV/nucleon. The red lines represent eq.~(\ref{Lo}) for masses from 1 to 12.}
\end{figure}

Figure~\ref{fig:MASS} shows an example of $\rho Z \sin(\theta)$ versus $\beta$ for a beam energy of 100~MeV/nucleon. The red curves represent the eq.~(\ref{Lo}) with $A$ varying from 1 to 12.

\section{Results}

Results are shown as values averaged over all detected type of fragments except in the case of the charge identification efficiencies.
However, to show the dependence of the results for different fragments, data for the trajectory reconstruction and mass identification efficiencies have also been extracted for protons, $^4$He, $^{11}$B and $^{11}$C using the optimal system configuration found. Those fragments represent approximately 5\%, 57\%, 6\% and 6\% (slightly depending on the beam energy) of the total detected fragments.

\subsection{Charge identification efficiency}
The charge identification efficiency is evaluated as the ratio between the number of fragments that have their charge correctly identified and the number of fragments that hit the TOF-wall. It mostly relates on the TOF system through the energy resolution on the released energy and the coincidence resolving time of the TOF. The trackers might affect it when the fragments encounter scattering in their material but their positions and spatial resolutions do not affect those results.

\begin{figure}[!ht]
 \begin{center}
    \includegraphics[width=0.99\linewidth]{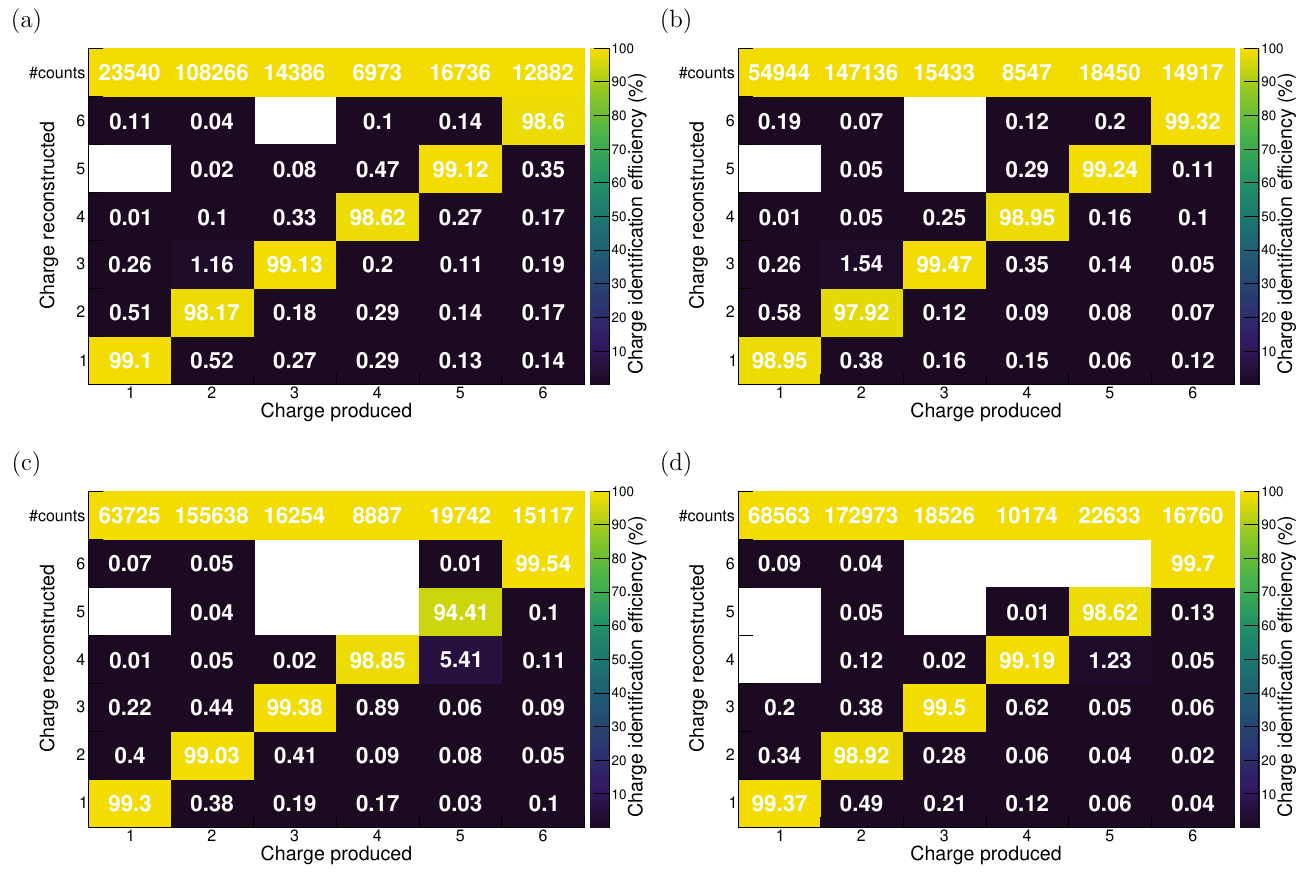}
 \end{center}
 \caption{\label{fig:CHARGE}Charge identification efficiency of the fragments for (a)~100~MeV/nucleon, (b)~200~MeV/nucleon, (c)~300~MeV/nucleon and (d)~400~MeV/nucleon.}
\end{figure}

Figure~\ref{fig:CHARGE} shows the charge identification efficiency matrices of the fragments for each beam energy. For most of the particle charges and beam energies, the identification is achieved with an efficiency above 98\%, reaching more than 99\% in some cases.


\subsection{Trajectory reconstruction efficiency}

Considering the way the algorithm works, the two main parameters that affects the trajectory reconstruction efficiency are the position of the detectors and their spatial resolutions.

The trajectory reconstruction efficiency is defined here as the ratio between the number of fragments that have their trajectory reconstructed and the number of fragments having a trajectory that went through all the detectors, hence reaching the TOF-wall detection modules.

\subsubsection{Tracker positions}

\begin{figure}[!ht]
 \begin{center}
    \includegraphics[width=0.99\linewidth]{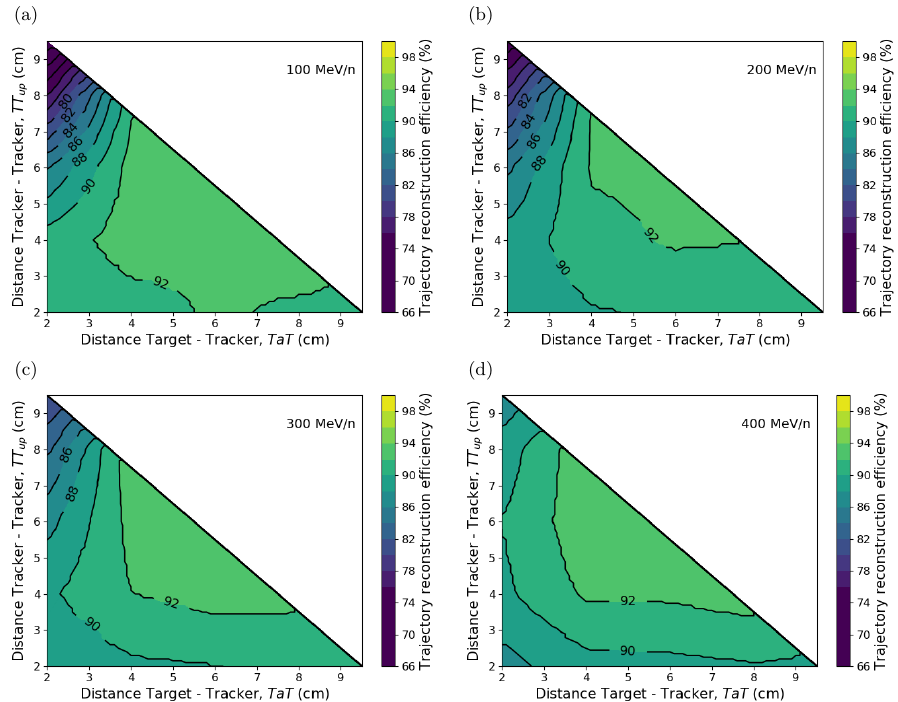}
 \end{center}
 \caption{\label{fig:EFFTRAJPOSUP}Trajectory reconstruction efficiency as a function of the upstream tracker positions at (a)~100~MeV/nucleon, (b)~200~MeV/nucleon, (c)~300~MeV/nucleon and (d)~400~MeV/nucleon.}
\end{figure}

Figure~\ref{fig:EFFTRAJPOSUP} shows the trajectory reconstruction efficiency as a function of the positions of the upstream trackers for the different beam energies.

For all the beam energies, the trajectory reconstruction efficiency is better than 91\% if the distance $TaT$ is greater than 4~cm and the distance between the two trackers $TT_{up}$ is larger than 4~cm. Placing the first upstream tracker in a shorter distance to the target and the two trackers far from each other is critical as the trajectory reconstruction efficiency drops significantly with smaller values of $TaT$, especially at 100 and 200~MeV/nucleon. Overall, the highest efficiencies are achieved when $4 \leq TaT \leq 8$~cm and $4 \leq TT_{up} \leq 7$~cm.

\begin{figure}[!ht]
 \begin{center}
    \includegraphics[width=0.99\linewidth]{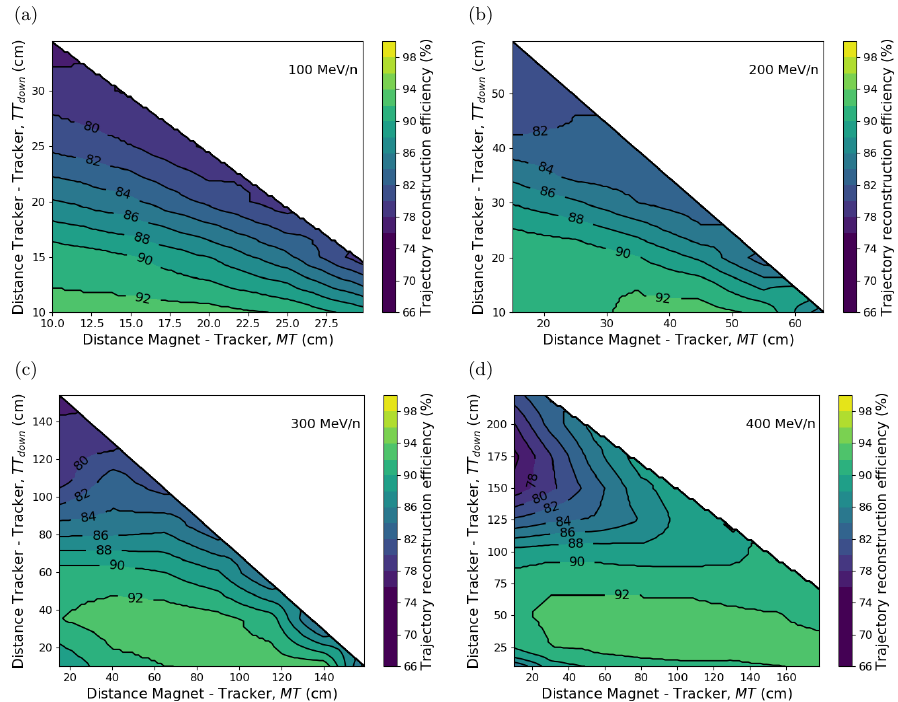}
 \end{center}
 \caption{\label{fig:EFFTRAJPOSDOWN}Trajectory reconstruction efficiency as a function of the downstream tracker positions at (a)~100~MeV/nucleon, (b)~200~MeV/nucleon, (c)~300~MeV/nucleon and (d)~400~MeV/nucleon.}
\end{figure}

Figure~\ref{fig:EFFTRAJPOSDOWN} shows the trajectory reconstruction efficiency as a function of the positions of the downstream trackers at the different beam energies.

The beam energy has a large influence on $MT$ and $TT_{down}$ to achieve a good trajectory reconstruction efficiency. In fact, to keep the efficiency above 92\% when increasing the beam energy, the lowest value of $MT$ should go from 10~cm at 100~MeV/nucleon to 40~cm at 400~MeV/nucleon. In the mean time, the highest values of $TT_{down}$ went from 12~cm to almost 70~cm.

\subsubsection{Tracker spatial resolutions}
In this part only the results at 400~MeV/nucleon are shown as the influence of the tracker spatial resolutions for the other beam energies were comparable.

The spatial resolution of the downstream trackers in the $x$ direction (i.e. the deflecting direction) have no clear influence on the trajectory reconstruction efficiency. In fact, it is stable at 92\% for a spatial resolution going from 100~\textmu m to 2500~\textmu m. This is mostly due to the fact that the spatial resolution of the downstream trackers in the $x$ direction is only used for the first step of the trajectory reconstruction, from the TOF-wall and the second downstream tracker to the first downstream tracker.

\begin{figure}[!ht]
    \begin{center}
        \includegraphics[width=0.5\linewidth]{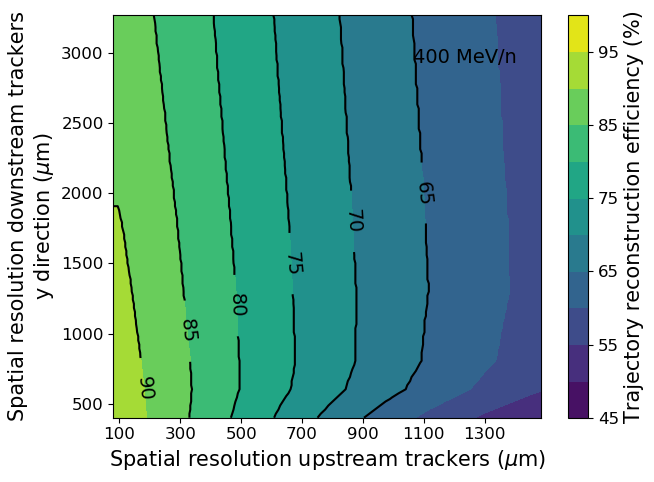}
    \end{center}
    \caption{\label{fig:EFFTRAJRES2}Trajectory reconstruction efficiency as a function of the upstream tracker spatial resolution and the downstream trackers spatial resolution in the $y$ direction at 400~MeV/nucleon. Results are averaged over all fragments are they mainly constrained by the lighter fragments.}
\end{figure}

Figure~\ref{fig:EFFTRAJRES2} shows the trajectory reconstruction efficiency as a function of the spatial resolution of the upstream trackers and the spatial resolution downstream trackers in the $y$ direction. Degrading the upstream spatial resolution from 100~\textmu m to 1500~\textmu m induces a loss of 35\% of efficiency. The optimal spatial resolution for the upstream trackers seems to be 100~\textmu m. For the downstream trackers, the loss of efficiency induced by degrading the spatial resolution in the $y$ direction from 500~\textmu m to 3000~\textmu m is less than 5\%. Setting it around 1~mm must be sufficient as lowering it would not give a better efficiency.
 
\subsection{Mass identification efficiency}
As for the trajectory reconstruction, the two parameters that influence the mass identification efficiency are also the positions of the trackers and their spatial resolutions. The mass identification efficiency is evaluated as the ratio between correctly identified mass fragments and the number of fragments that reached the TOF-wall. To evaluate the efficiency of the mass identification only, the algorithm used the charges and the trajectories given by the MC simulations.

\subsubsection{Tracker positions}
\begin{figure}[!ht]
 \begin{center}
    \includegraphics[width=0.99\linewidth]{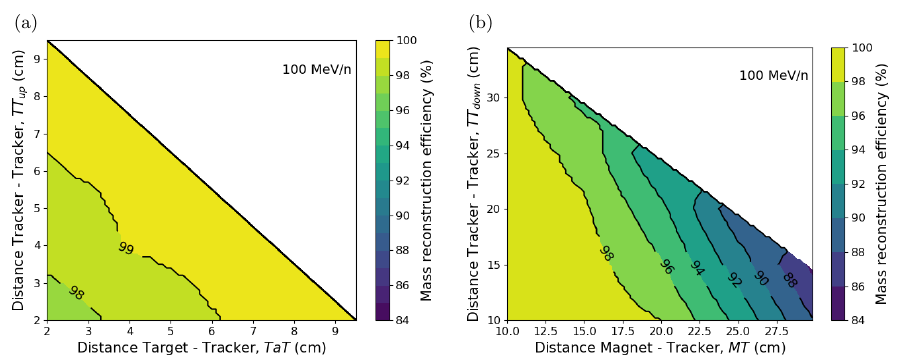}
 \end{center}
 \caption{\label{fig:EFFMASSPO}Mass identification efficiency as a function of the position of (a)~the upstream trackers, (b)~the downstream trackers at 100~MeV/nucleon.}
\end{figure}

Here only the results at 100~MeV/nucleon are shown as they gave results that are more constraining than the other beam energies, especially for the downstream tracker positions. Figure~\ref{fig:EFFMASSPO} shows the mass identification efficiency as a function of the positions of the upstream and the downstream trackers at 100~MeV/nucleon.

The positions of the upstream trackers show no significant influence on the mass identification efficiency. It is always kept above 97\% and the highest efficiencies are obtained when the second upstream tracker is the closest possible to the entry of the magnet.

The positions of the downstream trackers are however more critical parameters as the mass identification efficiency could fall down to 88\% in the worst case where the trackers are closer than 17~cm to each other. The mass identification efficiency is above 96\% when the distance between the exit of the magnet and the first downstream tracker $MT$ is kept below 17~cm and the distance between the trackers is lower than 20~cm.

\subsubsection{Tracker spatial resolutions}
\begin{figure}[!ht]
 \begin{center}
    \includegraphics[width=0.99\linewidth]{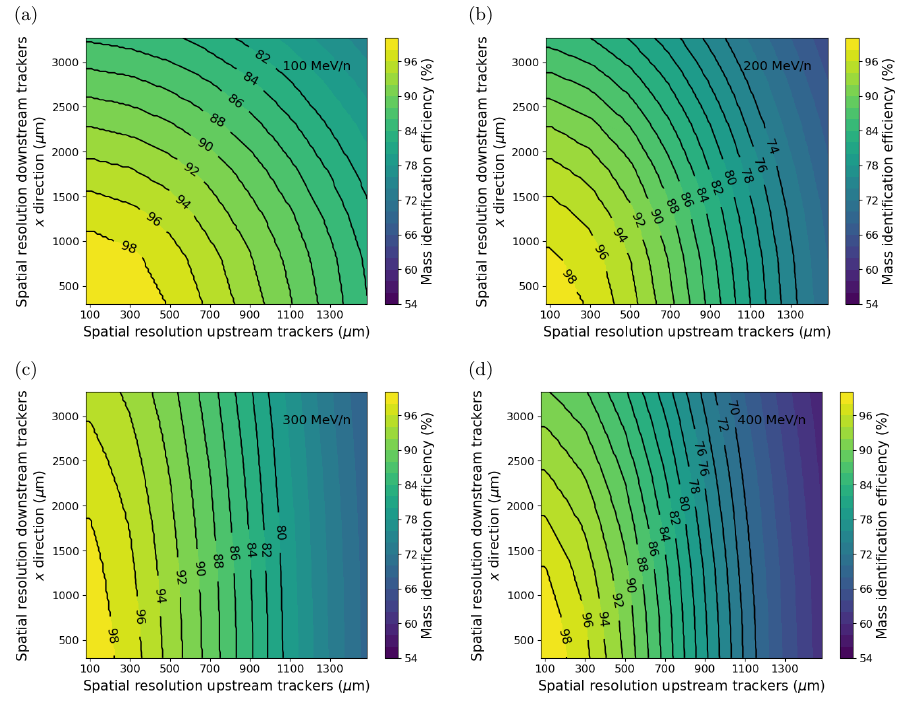}
 \end{center}
 \caption{\label{fig:EFFMASSRESX}Mass identification efficiency as a function of the spatial resolutions in the $x$ direction of the downstream trackers for different upstream tracker spatial resolutions at (a)~100~MeV/nucleon, (b)~200~MeV/nucleon, (c)~300~MeV/nucleon and (d)~400~MeV/nucleon.}
\end{figure}

Figure~\ref{fig:EFFMASSRESX} shows the mass identification efficiency as a function of the downstream trackers spatial resolutions in the $x$ direction for different upstream tracker spatial resolutions at the different beam energies.

The mass identification efficiency shows a strong dependency on the spatial resolution of the upstream trackers. For example at 100~MeV/nucleon and a downstream trackers $x$ spatial resolution of 1~mm, degrading the spatial resolution of the upstream trackers from 100~\textmu m to 1~mm reduces the efficiency from 96\% to less than 90\%. The effect is even more emphasised when increasing the beam energy with an efficiency around 74\% at 400~MeV/nucleon.

The spatial resolution of the downstream trackers has however a smaller effect on the mass identification than the upstream one. Degrading the $x$ spatial resolution from 1~mm to 3~mm results in a roughly 10\% loss of efficiency for the lowest beam energy to only few percents when increasing the energy. An optimal value of 1~mm for the spatial resolution of the downstream trackers in the $x$ direction was chosen as a better one would not improve drastically the mass identification efficiency.

\subsection{Optimal configurations}
Table~\ref{table:RESTRAJRECAP} summarizes the positions of the up and downstream trackers used in the optimal system configuration at the different beam energies deduced from the trajectory reconstruction and mass identification results. In these configurations, the upstream tracker spatial resolution was fixed at 100~\textmu m and the downstream tracker spatial resolutions at 1~mm and 1.5~mm in the $x$ and $y$ directions, respectively.
 
\begin{table}[!ht]
    \caption{\label{table:RESTRAJRECAP}Optimal positions of the trackers for the different beam energies.}
    \begin{center}
        \renewcommand{\arraystretch}{1.1}{
            \begin{tabularx}{0.85\linewidth}{XXXXX}
                \hline
                \hline
                { Beam energy (MeV/nucleon)} & { $TaT$ (cm) } & { $TT_{up}$ (cm) } & { $MT$ (cm) } & { $TT_{down}$ (cm) }  \\
				\hline
				\hline
				{ 100 }& { 6 } & { 4 } & { 15 } & { 10 }  \\
				{ 200 }& { 6 } & { 4 } & { 30 } & { 15 }  \\
				{ 300 }& { 6 } & { 4 } & { 65 } & { 35 } \\
				{ 400 }& { 6 } & { 4 } & { 105 } & { 25 }  \\
              \hline
              \hline
            \end{tabularx}}
        \end{center}
\end{table}

\begin{figure}[!ht]
 \begin{center}
    \includegraphics[width=0.99\linewidth]{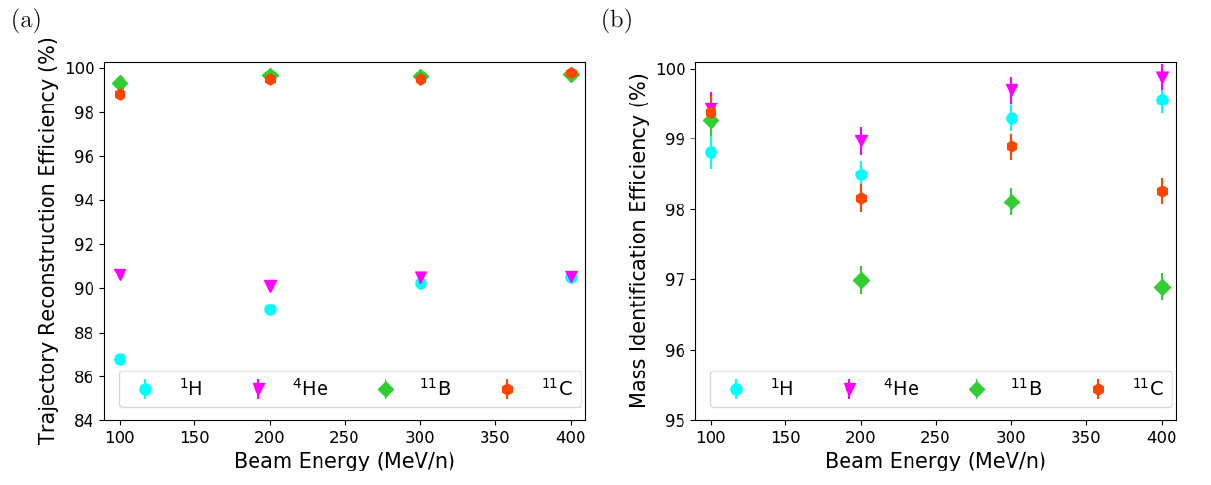}
 \end{center}
 \caption{\label{fig:TrajEffReco}(a) Trajectory reconstruction and (b) mass identification efficiencies for protons, $^4$He, $^{11}$B and $^{11}$C at the four different beam energies in the optimal configuration.}
\end{figure}

Figure~\ref{fig:TrajEffReco} shows the trajectory reconstruction and the mass identification efficiencies for protons, $^4$He, $^{11}$B and $^{11}$C at the four different beam energies in the optimal configuration. The trajectory reconstruction efficiency is clearly driven by the lightest fragments as the efficiencies for protons and $^4$He are lower or around 90\% while the results for the $^{11}$B and $^{11}$C are always above 98\% for all the beam energies. This is largely attributed to the small transverse momenta of the light fragments having very large magnetic rigidity. A small error on the up and downstream trajectories can lead to a wrong reconstructed radius. As shown in figure~\ref{fig:TrajEffReco}(b), when the trajectory is well reconstructed, the mass identification does not show this behaviour. A small decrease in the mass identification efficiencies can however be observed for $^{11}$B at 200 and 400~MeV/nucleon and for $^{11}$C at 400~MeV/nucleon. This has been attributed to a small but significant systematic uncertainty due to slight shifts of the A=11 identification lines towards the A=12 mass, overestimating the number of A=10 events.

\subsection{Fragment identification efficiencies}

\begin{figure}[!ht]
    \begin{center}
		\includegraphics[width=0.99\linewidth]{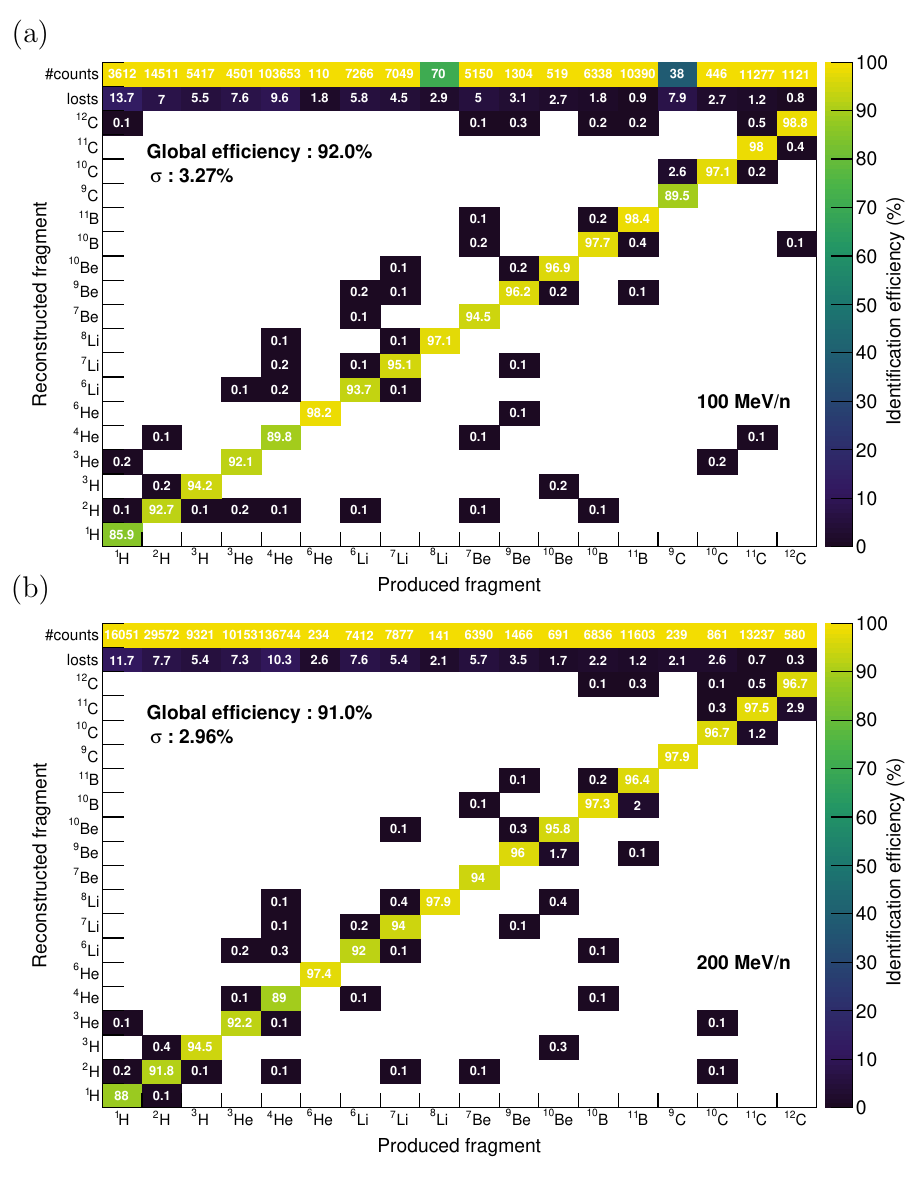}
    \end{center}
    \caption{\label{fig:EFFPARTID1}Fragment identification matrices for an optimal configuration of the apparatus at (a)~100~MeV/nucleon and (b)~200~MeV/nucleon.}
\end{figure}

\begin{figure}[!ht]
    \begin{center}
		\includegraphics[width=0.99\linewidth]{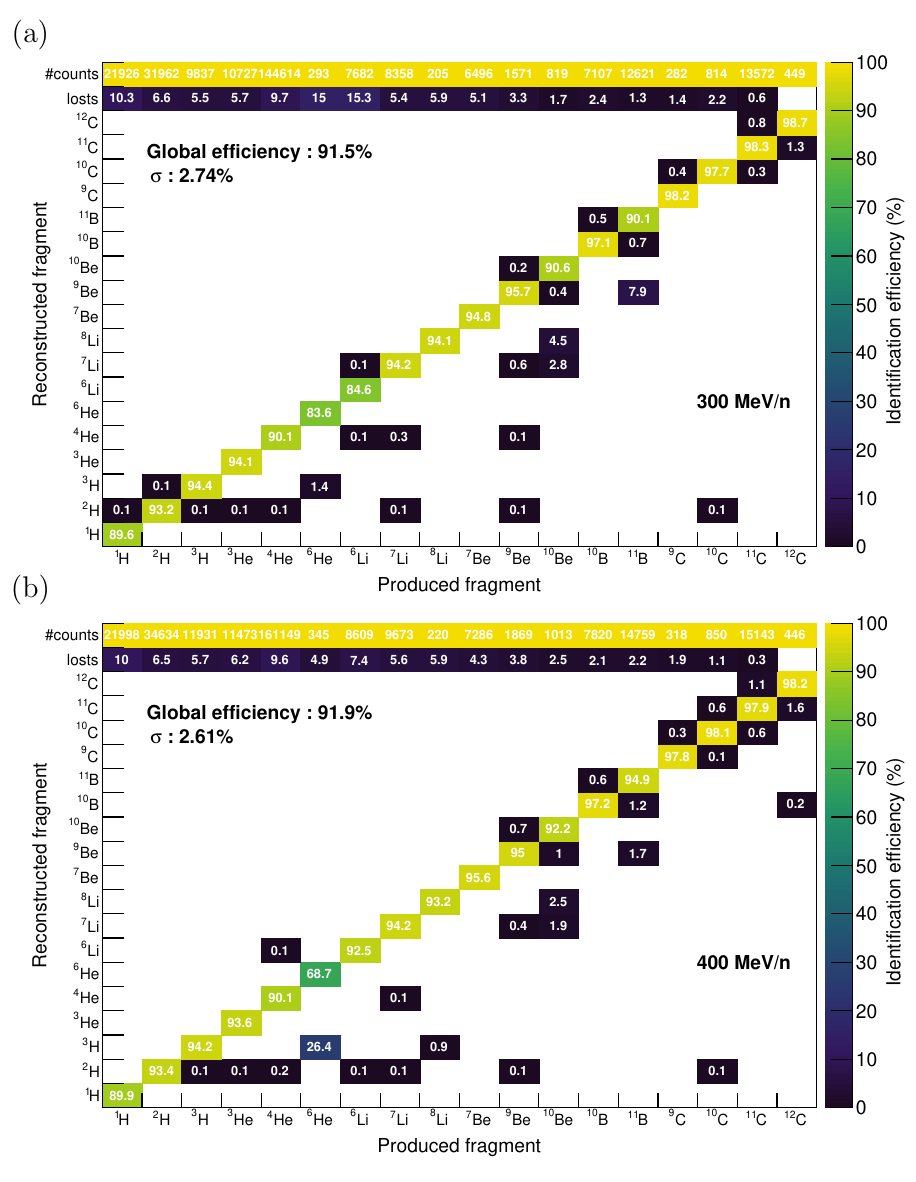}
	\end{center}
    \caption{\label{fig:EFFPARTID2}Fragment identification matrices for an optimal configuration of the apparatus at (a)~ 300~MeV/nucleon and (b)~400~MeV/nucleon.}
\end{figure}

Figure~\ref{fig:EFFPARTID1} and \ref{fig:EFFPARTID2} show the fragment identification matrices obtained with an optimal configuration of the apparatus given in table~\ref{table:RESTRAJRECAP}, at the different beam energies. These are made by comparing the number of identified fragments of a certain charge and mass (obtained after the charge identification, the trajectory reconstruction and the mass identification) to the one generated by the simulation. The mean and sigma values of the identification efficiencies are then calculated for all the fragments correctly identified. Globally, the performances are satisfactory as most of the fragments are correctly identified with an overall efficiency better than (90$\pm$3)\% for each beam energy. A significant proportion of the low fragment identification efficiencies are due to a wrong charge identification. For example at 300~MeV/nucleon the $^{10}$Be was identified for 5\% as $^{8}$Li and for 3\% as $^{7}$Li which drops its identification efficiency down to 91\%. Yet most of those low identification efficiencies concern fragments with a low production rate, while, on the opposite, the most produced fragments are globally well identified. For example, at each 4 energies, the $\alpha$-particles which represent roughly 50\% of the produced fragments have an identification efficiency above 89\%. 

The lost fragments correspond to fragments which reconstructed charges could not be associated to a mass due to a trajectory reconstructed with a different magnetic rigidity. The highest proportion of lost fragments is for protons which goes from almost 14\% at 100~MeV/nucleon to 10\% at 400~MeV/nucleon and can directly be related to the results given in figure~\ref{fig:TrajEffReco}(a). The lowest masses are again the most affected due to their smaller transverse momenta whereas for the other fragments, the proportion of losts is always under 10\%.

\section{Discussion}

The results given by the analysis of the Geant4 simulations of the FRACAS apparatus allow us to draw some conclusions concerning the characteristics needed for the up and downstream trackers. 

\subsection{Upstream trackers}

The results showed that at each beam energy the configuration giving the best trajectory reconstruction efficiency is by placing the first upstream tracker at 6~cm to the target and the second upstream tracker at 4~cm of the first. Given the fact that the mass reconstruction efficiency was not clearly affected by the position of the upstream trackers, we conclude that this is an optimal configuration of the positions of the upstream tracker for all beam energies. Placing the upstream trackers further away from the target or from each other will not lower the trajectory reconstruction efficiency but will decrease the geometric efficiency of the apparatus, unless the size of their active area was increased.

The trajectory reconstruction efficiency and the mass reconstruction efficiency both seemed to strongly rely on the spatial resolution of the upstream trackers. Thus, it will be crucial that the technology chosen for them provides a spatial resolution better than 100~\textmu m, while keeping the material budget as low as possible. As a solution, pixelated or stripped silicon detectors have shown to have a spatial resolution that can reach up to 2~\textmu m~\cite{AMBROSI1999215,HOU1999169} Another solution could be diamond detectors with stripped metallized anodes as they can provide a spatial resolution of around 26~\textmu m~\cite{BORCHELT1995318} and sustain a higher integrated flux without being damaged. However, both solutions being based on solid state detectors, they may increase the material budget above the requirements. 

\subsection{Downstream trackers}

Unlike the upstream trackers, the results concerning the downstream trackers showed that their positions must be changed for each beam energy. The trajectory reconstruction efficiencies and the mass reconstruction efficiencies were both mainly affected by the spatial resolution in the $x$ direction of the downstream trackers. The spatial resolution in the $y$ direction also influenced the mass reconstruction efficiency but at a lower level. 

Located after the mass separation of the large acceptance magnet, the downstream trackers should have a large active area to account for the very different particle momenta. The simplest and least expensive technology would then be to use gaseous detectors. This solution also allows to keep the material budget as low as possible. MWPCs can reach a spatial resolution of around 60~\textmu m in the direction along the wires and 200~\textmu m in the direction perpendicular to the wires~\cite{CHARPAK1979455}. Some studies are ongoing on a prototype of an MWPC to determine the spatial resolution reachable in both directions. We also plan to study the use of \textmu-RWELL~\cite{Bencivenni_2015} for the downstream trackers as they can reach a spatial resolution of around 50~\textmu m.

\subsection{Limitations of the study}

As the technologies for the different trackers were not introduced in the simulations, the results may be considered as ideal values and only show the effects of the tracker positions and spatial resolutions on the performances. Hence, even though the average multiplicity per event of the fragments on the up and downstream trackers were 4$\pm$1 and 3$\pm$1, respectively, the hit positions were correctly extracted from the simulations. The only fluctuation introduced was during the study of the spatial resolution. Furthermore, random hits due to the background noise or ghost hits related to the signal readout type have not been generated in the trackers as stated in the introduction. From the readout point of view, the deadtime introduced by the electronics or certain solid state technologies such as CMOS based pixel detectors~\cite{HUGUO2010480} could not be taken into account. All these effects would have a negative impact on the performances that was not evaluated in this work and will be studied more precisely in the future.

\section{Conclusion}

Geant4 simulations and in-house developed reconstruction algorithms permitted a systematic study of the influence of the positions and spatial resolutions of the detectors on the trajectory and the mass identification efficiencies of the FRACAS mass spectrometer.
With an optimal configuration of the positions and spatial resolutions of the trackers, it was possible to achieve particle identification efficiencies above 90\% with simulation data for beam energies ranging from 100 to 400~MeV/nucleon. Results showed that the position of the up and downstream trackers mostly affects the trajectory reconstruction efficiencies and that an optimal configuration can be determined at each beam energy.

The spatial resolution of the upstream trackers is a crucial parameter that strongly influences both trajectory reconstruction and mass identification and should be kept around 100~\textmu m to ensure acceptable fragment identification performances. Silicon or diamond pixelated or stripped detectors seem to be viable solutions for the development of the upstream trackers. Concerning the downstream trackers, the spatial resolution in the $x$ direction seem to be a lot more crucial than the spatial resolution in the $y$ direction. An optimal value of the spatial resolution could be 1~mm in the $x$ direction and 1.5~mm in the $y$ direction. MWPC or \textmu-RWELL detectors using stripped read-out can be viable technologies.

\bibliography{fracas}

\end{document}